\documentstyle[12pt]{article}
\newcommand{\II}{{\cal I}}
\newcommand{\JJ}{{\cal J}}

\newcommand{\be}{\begin{equation}}
\newcommand{\ee}{\end{equation}}
\newcommand{\ben}{\begin{eqnarray}\displaystyle}
\newcommand{\een}{\end{eqnarray}}
\newcommand{\refb}[1]{(\ref{#1})}

\begin{document}

{}~ \hfill\vbox{\hbox{hep-th/9603113}\hbox{MRI-PHY/96-10}}\break

\vskip 3.5cm

\centerline{\large \bf Orbifolds of $M$-Theory and String Theory}

\vspace*{6.0ex}

\centerline{\large \rm Ashoke Sen\footnote{On leave of absence from Tata
Institute of Fundamental Research, Homi Bhabha Road, Bombay 400005, INDIA}
\footnote{E-mail: sen@mri.ernet.in, sen@theory.tifr.res.in}}

\vspace*{1.5ex}

\centerline{\large \it Mehta Research Institute of Mathematics}
 \centerline{\large \it and Mathematical Physics}

\centerline{\large \it 10 Kasturba Gandhi Marg, Allahabad 211002, INDIA}

\vspace*{4.5ex}

\centerline {\bf Abstract}

It is shown that many of the conjectured dualities involving orbifold 
compactification of $M$-theory follow from the known dualities 
involving $M$-theory and string theory in ten dimensions, and the
ansatz that orbifolding procedure commutes with the duality
transformation. This ansatz also leads to a new duality conjecture,
namely that $M$-theory compactified on $T^8/Z_2$ is dual to type I
string theory on $T^7$. In this case the `twisted sector states' in
$M$-theory live on sixteen membranes transverse to the internal
manifold.

\vfill \eject

\noindent{\bf Introduction:}
In recent past it has been realised that the moduli space of
string theories contains a special point, known as the $M$-theory,
whose low energy limit corresponds to the eleven dimensional
supergravity theory\cite{TOWNSEND,WIS,SCHWARZ,ASPIN}. Compactification
of $M$-theory on various orbifolds have been studied, and have been
conjectured to be dual to various known string theory 
compactifications\cite{HORWIT,MUKHI,WITTK3,MTHEORY,DABH,FTHEORY,
DMW,EW,SW,ALOK,ACHA}.
In this paper we shall try to get a systematic understanding of some of
these duality conjectures in terms of other duality conjectures
involving $M$-theory and ten dimensional string theories; namely
the equivalence of $M$-theory on $S^1$ and type IIA string theory,
and the self-duality of type IIB string theory.

The basic idea is an old one. Suppose two theories $A$ and $B$ are
equivalent, and suppose further that the theory $A$ has a symmetry
$\sigma_A$ which gets mapped to a symmetry $\sigma_B$ under the
equivalence relation. Then naively we would expect that the theory
$A$ modded out by the symmetry $\sigma_A$ will be equivalent to the
theory $B$ modded out by the symmetry $\sigma_B$. Unfortunately, this
naive expectation does not always work, and there are specific 
instances where this procedure gives nonsensical answer\cite{VAFAWIT}.
It works when the equivalence relation between the two theories
involves a $T$-duality transformation rather than a more general
$U$-duality transformation; it also works when the adiabatic
argument given in ref.\cite{VAFAWIT} can be applied. Nevertheless,
there are many interesting examples of dual pairs constructed by 
this (or closely related) method where neither of these conditions
hold\cite{KACHRU,FHSV,VAFAWIT,FTHEORY,KS}. We shall start with the
assumption that this
procedure of obtaining dual pairs works even when one of the 
theories corresponds to $M$-theory compactification rather than
string theory compactification, and show that in most cases it
leads to sensible answers.\footnote{Some related observations were
made in ref.\cite{ACHA}.}
One of the exceptions is $M$-theory
compactification on $S^1/Z_2$, where this procedure would indicate
that the dual theory is type IIB string
theory, whereas in actual practice
it is known to be the $E_8\times E_8$ heterotic string 
theory\cite{HORWIT}. 

\noindent{\bf Notation:}
To begin with let us set up some notations. For either $M$-theory
or string theory compactified on an $n$ dimensional torus $T^n$,
we shall denote by $\II_n$ the transformation that changes the
sign of all the coordinates on the torus. This is usually not a
symmetry for odd $n$. For type IIA theory,
and for $M$-theory, this can be made into a symmetry by combining
it with an internal transformation, which, besides other effects,
changes the sign of the three form gauge field. We shall denote
this combined transformation by $\JJ_n$. For type IIA theory, 
the internal part of this transformation can be identified to
world-sheet parity transformation. For the type IIB theory, the 
world-sheet parity transformation is itself a symmetry of the
theory. We shall denote this transformation by $\Omega$. Finally,
both type IIA and type IIB theory possesses a symmetry which is
easiest to describe in the light-cone gauge Green-Schwarz formalism.
In this formalism this transformation changes the sign of all the 
left-moving fermions on the world-sheet. We shall denote this
transformation by $(-1)^{F_L}$, where $F_L$ stands for the space-time
fermion number arising in the left-moving sector of the world-sheet.
Acting on the massless bosonic fields in the theory, the effect of this
transformation is to change the sign of all the fields arising from the
Ramond-Ramond (RR) sector.

The only other $Z_2$ symmetry that we shall encounter in this paper is
a $Z_2$ symmetry transformation acting on a special class of $K3$
surfaces\cite{NIKULIN,SCHSEN,CHLO,MTHEORY}. This has the following
properties:

\begin{enumerate}

\item{Modding out the $K3$ surface by this symmetry
preserves $SU(2)$ holonomy.}

\item{Acting on the lattice of second cohomology elements of $K3$,
it exchanges the two $E_8$ factors leaving the rest of the lattice
invariant.}

\item{It has eight fixed points.}

\end{enumerate}

\noindent We shall denote this symmetry by $\sigma$.

We begin our discussion with $M$-theory on $S^1/\{1,\JJ_1\}$. According
to ref.\cite{HORWIT} this theory is equivalent to $E_8\times E_8$ 
heterotic string theory. On the other hand, by using the known duality
transformation between $M$-theory on $S^1$ and type IIA string theory,
one can easily verify that the transformation $\JJ_1$ in $M$-theory
gets mapped to the transformation $(-1)^{F_L}$ in the type IIA theory.
Thus naively we would conclude that $M$-theory on $S^1/\{1,\JJ_1\}$
should be equivalent to type IIA theory modded out by $\{1, 
(-1)^{F_L}\}$. However, the latter theory is known to be equivalent to 
the type IIB theory. Thus we see that the naive procedure of getting
dual pairs through orbifolding breaks down in this case. However, as
we shall see, this procedure does work in most of the other cases.

\noindent{\bf $M$-theory on $T^5/\{1, \JJ_5\}$:}
$M$-theory on $T^5$ is equivalent to type IIA on $T^4$.
In $M$-theory the transformation $\JJ_5$ is equivalent to $\JJ_1\cdot
\II_4$. This goes over to the symmetry $(-1)^{F_L}\cdot \II_4$ in
the type IIA theory. Thus we would conclude that $M$-theory on
$T^5/\{1, \JJ_5\}$ is equivalent to the type IIA theory on
$T^4/\{1, (-1)^{F_L}\cdot \II_4\}$. It is clear that the spectrum of
massless states coming from the untwisted sector in the two
theories will be identical, since by construction it is guaranteed
that the fields that are even (odd) in $M$-theory on $T^5$ under
$\JJ_5$ get mapped to fields that are even (odd) in type IIA on
$T^4$ under $(-1)^{F_L}\cdot \II_4$. Thus we only need to verify that
the spectra from the `twisted sector' agree. In $M$-theory the twisted
sector contributes 16 tensor multiplets\cite{MUKHI,WITTK3}. 
On the type IIA side, the
space-time part $\II_4$ of the $Z_2$ symmetry has sixteen fixed
points on $T^4$. Thus in order that the spectrum of massless states
in the two theories agree, one must verify that at each fixed point,
the twisted sector of the type IIA theory contains a single tensor
multiplet.

It is easy to verify this result by working in the light cone
gauge Green-Schwarz
formalism. First of all, one can analyse the surviving supersymmetry
generators, and verify that they belong to the chiral $N=2$ 
supersymmetry algebra in six dimensions, as is the case for 
$M$-theory on $T^5/\{1, \JJ_5\}$. The spectrum may be analysed as
follows.  Both on the left and the right hand sector there are
four periodic bosons and fermions and four anti-periodic bosons
and fermions. Thus the total vacuum energy vanishes. Quantization
of the eight fermionic zero modes (four from the left and four
from the right) gives a sixteen fold degenerate state. Through
careful analysis of the transformation laws of the 
fermion zero modes under the six dimensional Lorentz group
one can verify that 
there sixteen states indeed belong to a tensor multiplet of the
chiral $N=2$ supersymmetry algebra.

Instead of presenting the details of this calculation, we shall
map this type IIA orbifold into a type IIB orbifold via a
$T$-duality transformation, which would make this equivalence
obvious. Let us denote by $6,7,8,9$ the compact directions, and
make an $R\to 1/R$ duality transformation on the 6th coordinate. If 
$X^m_L$, $X^m_R$ ($6\le m\le 9$)
denote the left and right moving components of the bosonic
coordinates in the type IIA theory, and 
$Y^m_L$, $Y^m_R$ denote the left and right moving components of 
the bosonic coordinates in the type IIB theory, then the two sets
of variables are related as
\be \label{e1}
Y^6_L=-X^6_L, \qquad \qquad Y^m_L=X^m_L \quad \hbox{for} \quad
7\le m\le 9, \qquad \qquad Y^m_R=X^m_R \quad \hbox{for} \quad
6\le m \le 9\, .
\ee
Now both the transformations $\II_4$, as well as $(-1)^{F_L}$, in
the type IIA theory can be represented as a $T$-duality rotation 
of the form\footnote{This notation is similar to the one used in
\cite{SENVAFA}.} 
\ben \label{e2}
&& (X^6_L+iX^7_L, X^8_L+iX^9_L, X^6_R+iX^7_R, X^8_R+iX^9_R)
\nonumber \\
&\to&
(e^{i\theta_L}(X^6_L+iX^7_L), e^{i\phi_L} (X^8_L+iX^9_L), 
e^{i\theta_R}(X^6_R+iX^7_R), e^{i\phi_R}(X^8_R+iX^9_R))\, .
\een
In particular the transformation $\II_4$ corresponds to
\be \label{e3}
(\theta_L, \phi_L; \theta_R, \phi_R) = (-\pi, \pi; -\pi, \pi)\, ,
\ee
and the transformation $(-1)^{F_L}$ corresponds to
\be \label{e4}
(\theta_L, \phi_L; \theta_R, \phi_R) = (2\pi, 0; 0, 0)\, .
\ee
Note that the rotation by $2\pi$ has no effect on the bosonic
coordinates which transform as vectors, but acts on the left
moving fermions transforming in the spinor representation
as a change of sign. This is precisely the effect of $(-1)^{F_L}$.
Thus the combined effect of $(-1)^{F_L}$ and $\II_4$ is given by
\be \label{e5}
(\theta_L, \phi_L; \theta_R, \phi_R) = (\pi, \pi; -\pi, \pi)\, ,
\ee
{}From eq.\refb{e1} we see that the transformation \refb{e2} can
be rewritten in terms of the coordinates in the type IIB theory
as
\ben \label{e6}
&& (Y^6_L+iY^7_L, Y^8_L+iY^9_L, Y^6_R+iY^7_R, Y^8_R+iY^9_R)
\nonumber \\
&\to&
(e^{i\theta_L'}(Y^6_L+iY^7_L), e^{i\phi_L'} (Y^8_L+iY^9_L), 
e^{i\theta_R'}(Y^6_R+iY^7_R), e^{i\phi_R'}(Y^8_R+iY^9_R))\, ,
\een
where
\be \label{e7}
(\theta_L',\phi_L';\theta_R', \phi_R')
= (-\theta_L,\phi_L;\theta_R, \phi_R)\, .
\ee
{}From eqs.\refb{e5} and \refb{e7}
we see that the transformation $(-1)^{F_L}\cdot \II_4$ in
the type IIA theory corresponds to
\be \label{e8}
(\theta_L',\phi_L';\theta_R', \phi_R')
=(-\pi, \pi; -\pi, \pi)
\ee
in the type IIB theory. This can easily be identified to be the
transformation $\II_4$ in type IIB theory.\footnote{One can also verify
explicitly, by using the standard map between the massless fields
in the type IIA and the type IIB theories under duality 
transformation, that the action of $(-1)^{F_L}\cdot\II_4$ on the
massless fields in the type IIA theory is identical to that of
$\II_4$ on the massless fields in the type IIB theory.}
This leads us to conclude
that type IIA on $T^4/\{1, (-1)^{F_L}\cdot \II_4\}$ is equivalent to
type IIB on $T^4/\{1, \II_4\}$. 

Using the earlier equivalence between $M$-theory on $T^5/\{1, \JJ_5\}$
and type IIA theory on
$T^4/\{1, (-1)^{F_L}\cdot \II_4\}$ we then conclude
that $M$-theory on $T^5/\{1, \JJ_5\}$ is equivalent to type IIB on
$T^4/\{1, \II_4\}$. The latter is a special case of type IIB theory on
a K3 surface, and is known to give
one tensor multiplet from each of the fixed
points in the twisted sector. In fact the duality between type IIB on
$K3$ and $M$-theory on $T^5/\{1, \JJ_5\}$ has already been conjectured 
in refs\cite{MUKHI,WITTK3}. Thus we have reproduced the conjectured 
duality of refs.\cite{MUKHI,WITTK3} by assuming that the orbifolding
procedure commutes with the duality transformation. 

\noindent{\bf $M$-theory on $T^9/\{1, \JJ_9\}$:}
This case may be analyzed along more or less similar lines as the
previous model. Using identical logic we see that this theory is
expected to be dual to type IIA theory on $T^8/\{1, (-1)^{F_L}
\cdot \II_8\}$. Under a further $R\to 1/R$ duality in one of the
eight compact directions, this theory reduces to type IIB on
$T^8/\{1, \II_8\}$. The duality between $M$-theory on $T^9/
\{1, \JJ_9\}$ and type IIB on $T^8/\{1, \II_8\}$ has already
been conjectured\cite{UNPUB}, where it was shown that each of
the 256 fixed points in the type IIB theory gives one left-moving
chiral boson neutral under supersymmetry,
which upon fermionization maps to 512 chiral fermions, one 
associated with each of the 512 fixed points of $T^9$ under
$\JJ_9$ in the $M$-theory.

\noindent{\bf $M$-theory on $(S^1 \times K3)/\{1, \JJ_1\cdot \sigma\}$:}
This model was analyzed in ref.\cite{MTHEORY}. The spectrum of massless
states in this model corresponds to that of
a chiral $N=1$ supergravity theory
with nine tensor-, eight vector- and twenty huper-multiplets. 
Using the duality
between $M$ theory on $S^1$ and the type IIA theory, and noting that the
symmetries $\JJ_1$ and $\sigma$ in $M$-theory corresponds to 
$(-1)^{F_L}$ and
$\sigma$ respectively in the type IIA theory, we can map this model to
type IIA on $T^4/\{1, (-1)^{F_L}\cdot \sigma\}$. The spectra of
massless states in the untwisted sector match trivially.
The spectra of massless states in the twisted sector can be seen to match in
the following way. In the $M$-theory orbifold, the `twisted sector
states' consist of eight tensor- and eight 
hyper-multiplets of the chiral $N=1$ supersymmetry 
algebra\cite{MTHEORY}, thus we
need to show that the twisted sector states in the type IIA orbifold
also consist of eight tensor multiplets. Since $K3/\{1, \sigma\}$ 
has eight fixed points, this amounts to proving that there is
one tensor- and one hyper- multiplet coming from each fixed point.
The spectrum of massless states coming from the twisted sector of
the type IIA orbifold can be computed by noting that locally the
space $K3/\{1, \sigma\}$ near the fixed point has the same structure 
as the space $T^4/\{1, \II_4\}$. Thus the spectrum of massless
states per fixed point from the twisted sector must be the same
as that in type IIA on $T^4/\{1, \II_4\}$. As has already been
argued, the latter theory has one tensor multiplet of the $N=2$
supersymmetry algebra per fixed point. This corresponds to
one tensor- and one hyper-multiplet of the $N=1$ supersymmetry 
algebra. Thus type IIA on $K3/\{1, (-1)^{F_L}\cdot \sigma\}$
does have eight hyper- and eight tensor- multiplets from the
twisted sector.\footnote{Incidentally, this
provides us with a new class of string compactification with more
than one tensor multiplets in six dimensions.}
This shows that the massless spectrum of $M$-theory
on $(S^1\times K3)/\{1, \JJ_1\cdot \sigma\}$ agrees with that of
type IIA string theory on 
$K3/\{1, (-1)^{F_L}\cdot \sigma\}$.

We would also like to relate this to a type IIB compactification.
Conventional $T$-duality transformation on type IIA on $K3$ will
give us back type IIA on $K3$, so we need to use a different
strategy. For this we shall choose a special $K3$ surface, namely
an orbifold $T^4/\{1, \II_4\}$. We do not suffer from
any loss of generality this way, since once we establish the
duality at one point in the moduli space, it holds at all other
points as well. Also for convenience we shall take the $T^4$ to
be the product of four circles, each at the self-dual radius.
Let $\eta$ denote a transformation
\be \label{e9}
(X^6, X^7, X^8, X^9) \to (X^6, X^7, X^8, X^9+\pi R)\, ,
\ee
where $X^m$ denote the coordinates on $T^4$ and $R$ denotes the 
radius of the ninth circle.\footnote{Note that $\eta$ commutes
with $(-1)^{F_L}$ and $\II_4$, since translation of $X^9$ by
$2\pi R$ is an identity transformation on $T^4$.}  Then if we
define our $K3$ surface to be $T^4/\{1,\II_4\}$, the transformation 
$\sigma$ on this $K3$ can be identified to 
$\II_4 \cdot \eta$.\footnote{This is a slightly different notation
from ref.\cite{DABH}.} There are sixteen fixed points of this
transformation on $T^4$, but modding out by $\II_4$ identifies them
pairwise. This gives eight fixed points on $K3$ as expected.

Thus our starting point, type IIA on $K3/\{1, 
(-1)^{F_L}\cdot \sigma\}$, is now
represented as type IIA on $T^4/\{1,\II_4,(-1)^{F_L}
\cdot \II_4\cdot\eta, (-1)^{F_L}\cdot \eta\}$.
The transformation $\II_4$, in the notation of eq.\refb{e2}, is
represented by the transformation given in \refb{e3}. Let us
now make an $R\to 1/R$ duality transformation in the sixth
coordinate. This would convert the type IIA theory to type IIB.
Using eqs.\refb{e3}, \refb{e4} and \refb{e7} we see that
the transformation $(-1)^{F_L}$ and $\II_4$ in the type IIA theory 
are mapped to $(-1)^{F_L}$  and $(-1)^{F_L}\cdot \II_4$
in the type IIB theory respectively. The transformation $\eta$
remains the same in the type IIB theory, as this represents a
shift in the 9th direction whereas the duality transformation is
being performed in the 6th direction. This leads us to the
conclusion that the theory under consideration is equivalent to
type IIB on $T^4/\{1, (-1)^{F_L}\cdot \II_4, \II_4\cdot \eta,
(-1)^{F_L}\cdot \eta\}$. Let us now trade in the coordinates $Y^m$
labelling the $T^4$ of the type IIB theory in favour of new 
coordinates $Z^m$ defined as
\be \label{e10}
Z^m=Y^m \qquad \hbox{for} \qquad 6\le m\le 8\, ,
\qquad \qquad Z^9=Y^9 + (\pi R/2)\, .
\ee
In terms of the new coordinates $Z^m$, the symmetries 
$\{1, (-1)^{F_L}\cdot \II_4, \II_4\cdot \eta,
(-1)^{F_L}\cdot \eta\}$ map onto 
$\{1, (-1)^{F_L}\cdot \II_4\cdot \eta, \II_4,
(-1)^{F_L}\cdot \eta\}$ with $\eta$ denoting shift of $Z^9$ by
$\pi R$. Noting that $T^4/\{1, \II_4\}$ denotes a $K3$ surface,
and furthermore, that the symmetry $\II_4\cdot\eta$ denotes the
transformation $\sigma$ on this $K3$ surface, we can identify the
model as type IIB  on $K3/\{1, (-1)^{F_L}\cdot \sigma\}$. 

Note that
during the course of the manipulation that we have performed in
going from the type IIA compactification to the type IIB 
compactification, the $Z_2$ transformation $\II_4$ in the type IIA
theory that was responsible for creating the $K3$ out of $T^4$
has been mapped to $\sigma\cdot (-1)^{F_L}$ acting on the $K3$
surface on which the type IIB theory is compactified. On the other
hand, the $Z_2$ transformation that generated $(-1)^{F_L}
\cdot \sigma$ on the
original $K3$ surface has been mapped to the $Z_2$ transformation that
generates the final $K3$ out of $T^4$. As a result, viewed as
compactification on an orbifold of $K3$, the twisted sector
states in the original theory will get mapped to the untwisted
sector states of the final theory. On the other hand, the
twisted sector states in the final theory will be mapped onto
the untwisted sector states of the original theory.

We can now map this model to another model by using the $S$-duality
of the type IIB theory in ten dimension\cite{HT}. Under this, the 
transformation $(-1)^{F_L}$ gets mapped to the world-sheet parity
transformation $\Omega$\cite{VAFAWIT}, whereas the transformation
$\sigma$ is unchanged. Thus one would expect that the
type IIB compactification that we have obtained is also equivalent
to type IIB on $K3/\{1, \Omega\cdot \sigma\}$. Note that this 
implicitly assumes that the $S$-duality transformation commutes 
with the orbifolding procedure, and since this $S$-duality is not
part of a $T$-duality group, we must explicitly check that the
spectrum from the twisted sector matches between the two theories.
However, the latter model is precisely the one that was analysed by 
Dabholkar and Park\cite{DABH} and was shown to have identical
spectrum of massless states as $M$-theory on $(S^1\times K3)/
\{1, \JJ_1\cdot\sigma\}$ at a generic point in the moduli space. 
Thus we see that following this chain
of dualities we have been able to `prove' the equivalence between
the $M$-theory compactification on $(S^1\times K3)/\{1, 
\JJ_1\cdot \sigma\}$ and type IIB compactification on 
$K3/\{1,\Omega\cdot\sigma\}$.
We have also seen that at one stage in this chain of dualities,
the twisted sector states and part of the untwisted sector states
got exchanged. This would explain why the eight tensor multiplets
that arise from the `twisted sector' of $M$-theory on 
$(S^1\times K3)/\{1, \JJ_1\cdot\sigma\}$ appear in the untwisted 
sector in type IIB on $K3/\{1,\Omega\cdot\sigma\}$, and the eight
vector multiplets that arise in the `twisted sector' in type
IIB on $K3/\{1, \Omega\cdot\sigma\}$ appear in the untwisted
sector in $M$-theory on $(S^1\times K3)/\{1, \JJ_1\cdot\sigma\}$.

\noindent{\bf $M$-theory on $T^4/\{1,\II_4\}$:}
This is a special case of $M$-theory on $K3$, and so we expect
the spectrum of massless states to be identical to that of $M$-theory
on $K3$. Using the relationship between the massless fields in
$M$-theory on $S^1$ and type IIA theory, we can easily identify
the symmetry $\II_4$ in $M$-theory on $T^4$ as the symmetry
$\JJ_3\cdot (-1)^{F_L}$ in the type IIA theory.\footnote{One
way to see this is to note that upon rewriting $T^4$ as
$S^1\times T^3$, the transformation $\II_4$ in
the $M$-theory can be regarded as the product $\JJ_1\cdot\JJ_3$.
$\JJ_1$ maps to $(-1)^{F_L}$ in the type IIA theory, whereas
$\JJ_3$ of $M$-theory is mapped to $\JJ_3$ of the type IIA theory.}
Thus this particular $M$-theory compactification
is expected to be dual to type IIA on $T^3/\{1,\JJ_3
\cdot (-1)^{F_L}\}$. This is an orientifold\cite{ORIENT} since $\JJ_3$
involves a world-sheet parity transformation.
One could analyse the spectrum of massless states in this theory
directly, but we can simplify the analysis by making a $T$-duality
transformation that inverts the radii of {\it all the three circles}
of $T^3$. This converts the type IIA theory to type IIB theory. 
Using the
standard relationship between the massless fields in the two theories
it is easy to see that the transformation $\JJ_3\cdot (-1)^{F_L}$
gets mapped to the world sheet parity transformation $\Omega$ in
the type IIB theory. Thus this orbifold compactification is
equivalent to type IIB on $T^3/\{1,\Omega\}$. This is nothing but the
type I string theory compactified on $T^3$. Thus we conclude
that $M$-theory on $T^4/\{1, \II_4\}$ is equivalent to type I theory on
$T^3$. Using the conjectured duality between type I and SO(32)
heterotic string theory in ten dimensions, this can be further
related to heterotic string theory compactified on $T^3$. The
equivalence between $M$-theory on $K3$ and heterotic string theory
on $T^3$ has already been conjectured in ref.\cite{WIS}.

\noindent{\bf $M$-theory on $T^8/\{1,\II_8\}$:}
The analysis in this case proceeds exactly as in the previous case,
leading us to the conclusion that this theory is equivalent to
type IIA on $T^7/\{1,\JJ_7\cdot (-1)^{F_L}\}$, which, in turn, is
equivalent to the type I theory on $T^7$. In the type I theory,
viewed as an orientifold of the type IIB theory, the `twisted sector
states' are open strings living on 32 nine-branes\cite{ORIENT} and
give rise to SO(32) gauge fields and their superpartners.
The duality transformation that maps this to the type IIA theory
on $T^7/\{1, (-1)^{F_L}\cdot\JJ_7\}$ involves inverting the radius
of each of the seven circles of $T^7$, and converts the 32
nine-branes into 32 membranes. At a generic point in the moduli
space these 32 membranes can be grouped into 16 pairs, the membranes
in each pair being related by the transformation $(-1)^{F_L}\cdot\JJ_7$.
Since the map from the type IIA theory to $M$-theory on $S^1$
converts a membrane to a membrane, we see that in the $M$-theory the
twisted sector states live on 16 membranes 
moving on the internal manifold $T^8/\{1, \II_8\}$ (which would appear as
32 membranes on $T^8$).\footnote{This, in
turn, implies that each of the 256 fixed points on $T^8/\{1, \II_8\}$
acts as a source of $-1/16$ units of anti-symmetric tensor field
charge.} 
Each of these 16 membranes contributes one
vector multiplet of the supersymmetry algebra.
This is analogous
to the situation for $M$ theory on $T^5/\{1,\JJ_5\}$, where the
`twisted sector states' come from the sixteen five-branes moving
on the internal manifold, each of which correspond to one
tensor multiplet of the chiral $N=2$ supersymmetry algebra.

\noindent{\bf Conclusion:}
We have seen that in many cases the duality involving $M$-theory
orbifolds may be understood by assuming that duality transformations
commute with orbifolding. There are, however, specific
examples where this procedure breaks down, notably when one gets
extra supersymmetry generators in the `twisted sector' of one of
the theories. It will be extremely interesting to have a systematic
understanding of when and why duality transformation commutes
with the orbifolding procedure, since this will give us a global
understanding of the interconnections between various dualities
in string theory.

I wish to thank K. Dasgupta and S. Mukhi for useful discussions.

\end{document}